\documentclass[twocolumn,10pt,aps,prl,preprint]{revtex4}   
\usepackage{amsmath}    
\usepackage{graphicx}   
\begin{document}
\draft
\newcommand{\ajp}{AJP}  

\title{Observing the evaporation transition in vibro-fluidized granular matter}
\author{Jorge E. Fiscina$^{1}$ and Manuel O. C\'{a}ceres$^{2,3,4}$}
\address{$^{1}$Experimental Physics-Saarland University, D-66123, Saarbr\"{u}cken Germany.\\$^{2}$Centro At\'{o}mico Bariloche, Instituto Balseiro, and $^{3}$CONICET,
\\8400 Bariloche, Argentina; and $^{4}$The Abdus Salam ICTP,34014 Trieste, Italy.}
\date{submitted for publication, September 1$^s$$^t$, 2006}

\begin{abstract}
By shaking a sand box the grains on the top start to jump giving
the picture of evaporating a sand bulk, and a gaseous transition
starts at the surface granular matter (GM) bed. Moreover the
mixture of the grains in the whole bed starts to move in a
cooperative way which is far away from a Brownian description. In
a previous work we have shown that the key element to describe the
statistics of this behavior is the exclusion of volume principle,
whereby the system obeys a Fermi configurational approach \cite
{PRL9510}. Even though the experiment involves an archetypal
non-equilibrium system, we succeeded in defining a global
temperature, $\beta ^{-1}$, as the quantity associated to the
Lagrange parameter in a maximum entropic statistical description.
In fact in order to close our approach we had to generalize the
equipartition theorem for dissipative systems. Therefore we
postulated, found and measured a fundamental dissipative parameter, $\delta $%
, written in terms of pumping and gravitational energies, linking
the configurational entropy to the collective response for the
expansion of the centre of mass (c.m.) of the granular bed. Here
we present a kinetic approach to describe the experimental
velocity distribution function (VDF) of this non-Maxwellian gas of
macroscopic Fermi-like particles ({\it mFp}). The evaporation
transition occurs mainly by jumping balls governed by the excluded
volume principle. Surprisingly in the whole range of low
temperatures that we measured this description reveals a
lattice-gas, leading to a packing factor, $p_{f}$, which is
independent of the external parameters. In addition we measure the
{\it mean free path,} $L$, as a function of the driving frequency,
$f_{e}$, and corroborate our prediction from the present kinetic
theory.
\end{abstract}
\maketitle

The granular matter bed was set up with $5437$ balls of $%
Z_{r}O_{2}-Y_{2}O_{3}$, in a glass container of $D_{c}=50$mm diameter (which
is open on the top) in a chamber at $1$atm of air under low water vapor
content conditions ($5.8\pm 0.2$gr/m$^{3}$), see Fig. 1(a,b). The diameter
of the balls was $D=1.99$mm and their mass $m=26.8\pm 0.1$mgr. Under
vertical vibration, it is possible to get the balls at the lower
gravitational position of the container in a crystalline state (the bulk),
and those at the upper one in a gas state, which is called the fluidized gap
\cite{PRL9510}. This non-Markovian gas doesn't expand to the whole volume of
the container, and the number of the {\it active} moving particles, in the
gas phase, is a function of the pumping energy. Thus $L$ of the {\it mFp}
determines the occupied volume of the evaporated gas.
\begin{figure}
  \includegraphics[width=8cm]{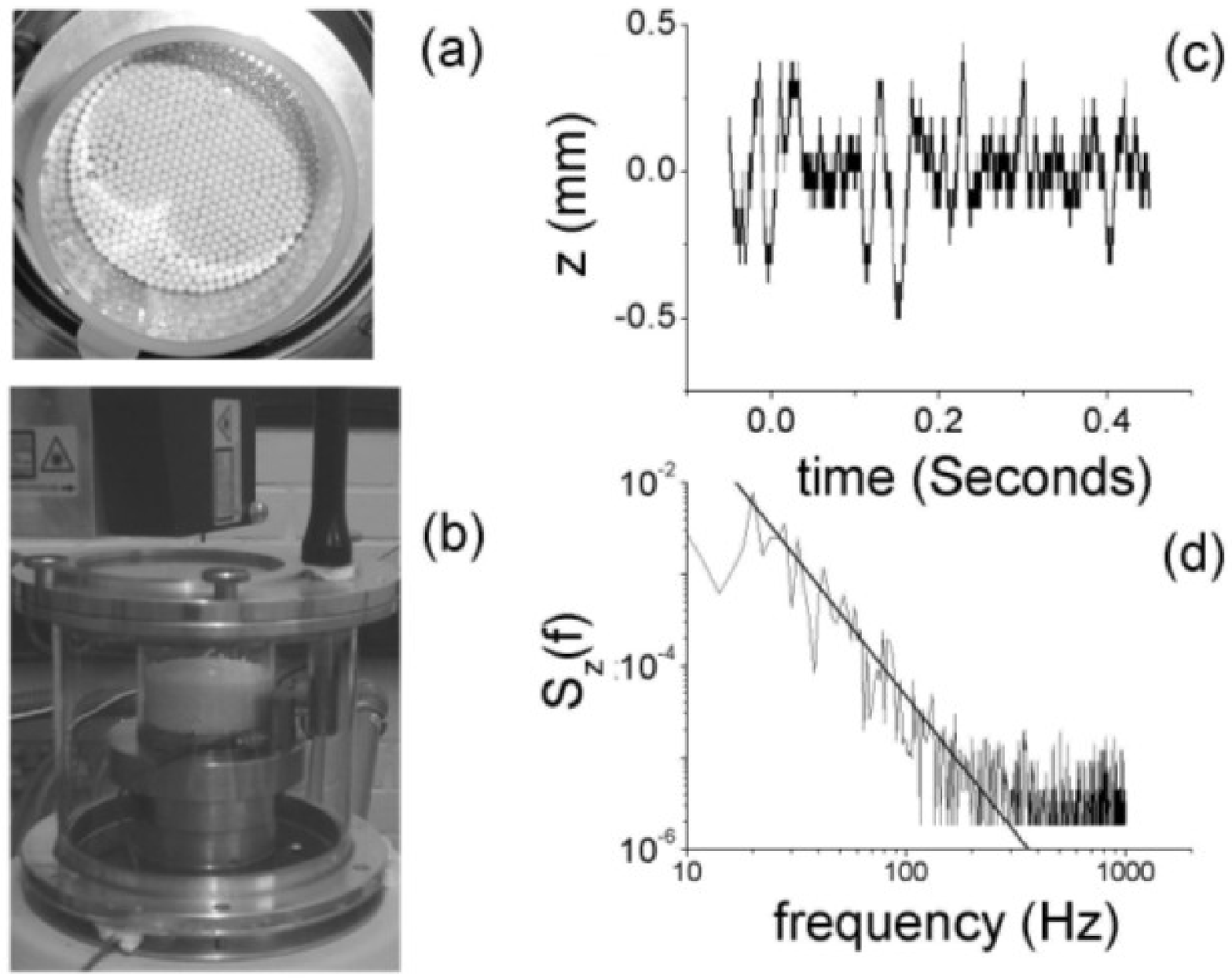}\\
  \caption{(a) Top view of the GM bed in a condensed fcc state. (b) Set up for
distance measurement (amplitude realizations $z(t)$) corresponding
to $11$ layer GM ($h=$ $18$mm) with $Z_{r}O_{2}-Y_{2}O_{3}$ balls
($D=1.99$mm) in a glass container of $D_{c}=50$mm diameter. (c)
Stochastic realization $z(t)$
at $83$Hz, $10$g, and (d) the corresponding anomalous power law $%
f^{-\upsilon }$ spectrum with $\upsilon =3$.}\label{FIG1:}
\end{figure}
A sinusoidal vibration is driven by a vibration plate on the GM bed of
height $h$ (with intensity $\Gamma =A\omega ^{2}/g$, where $A$ is the
amplitude, $g$ is the acceleration of gravity, and $\omega =2\pi f_{e}$).
The vibration apparatus is set up by an electromagnetic shaker (TIRAVIB$%
\,5212$) which allows for feedback through a piezoelectric accelerometer for
the control of $f_{e}$ and $\Gamma $ in the range of $10$-$7000$Hz, and $%
2-40g$ respectively. The control loop is completed by an Oscillator
Lab-works SC$121$ and a TIRA$\,19$/z amplifier of $1$kw. The vertical
trajectory (stochastic realization) $z(t)$ of one particle was followed in a
window of $12$mm with a laser device by using a triangulation method, see
Fig. 1(c) and its spectrum is shown in Fig. 1(d). A laser emitter with a
spot of $70\mu $m and a linear image sensor (CCD-like array) enables a high
speed measurement with $100\mu $sec sampling. The linear image sensing
method measures the peak position values for the light spots and suppresses
the perturbation of secondary peaks, which makes possible a resolution of $%
1\mu $m. The shaker and the laser displacement sensor were placed on
vibration-isolated tables to decouple them from external perturbations. $%
z(t) $ were taken with a $9354\,$C Le Croy Oscilloscope of $500$MHz. The
velocity, $V(t)=\left. dz\right/ dt$, of the {\it mFp} was calculated
numerically for $\Delta t=100\mu $sec from $z(t)$ registers. The amplitude
dispersion $\sigma _{z}=\sqrt{\left\langle z(t)^{2}\right\rangle
-\left\langle z(t)\right\rangle ^{2}}$ and velocity variance $\sigma
_{V}^{2}=$ $\left\langle V(t)^{2}\right\rangle -\left\langle
V(t)\right\rangle ^{2}$ were obtained from a window of $2$ seconds for each
pair of registers $\left\{ z(t),V(t)\right\} $. In Fig. 2(a) we report $%
\sigma _{z}$ against $f_{e}$ (from $40$-$100$Hz) for fixed $\Gamma =10$, for
a GM\ bed of $h=18$mm. We test experimentally in Fig.2(b) the relation
between $\sigma _{z}$ and $\sigma _{V}^{2}$.
\begin{figure}
  \includegraphics[width=5.68cm,angle=-90]{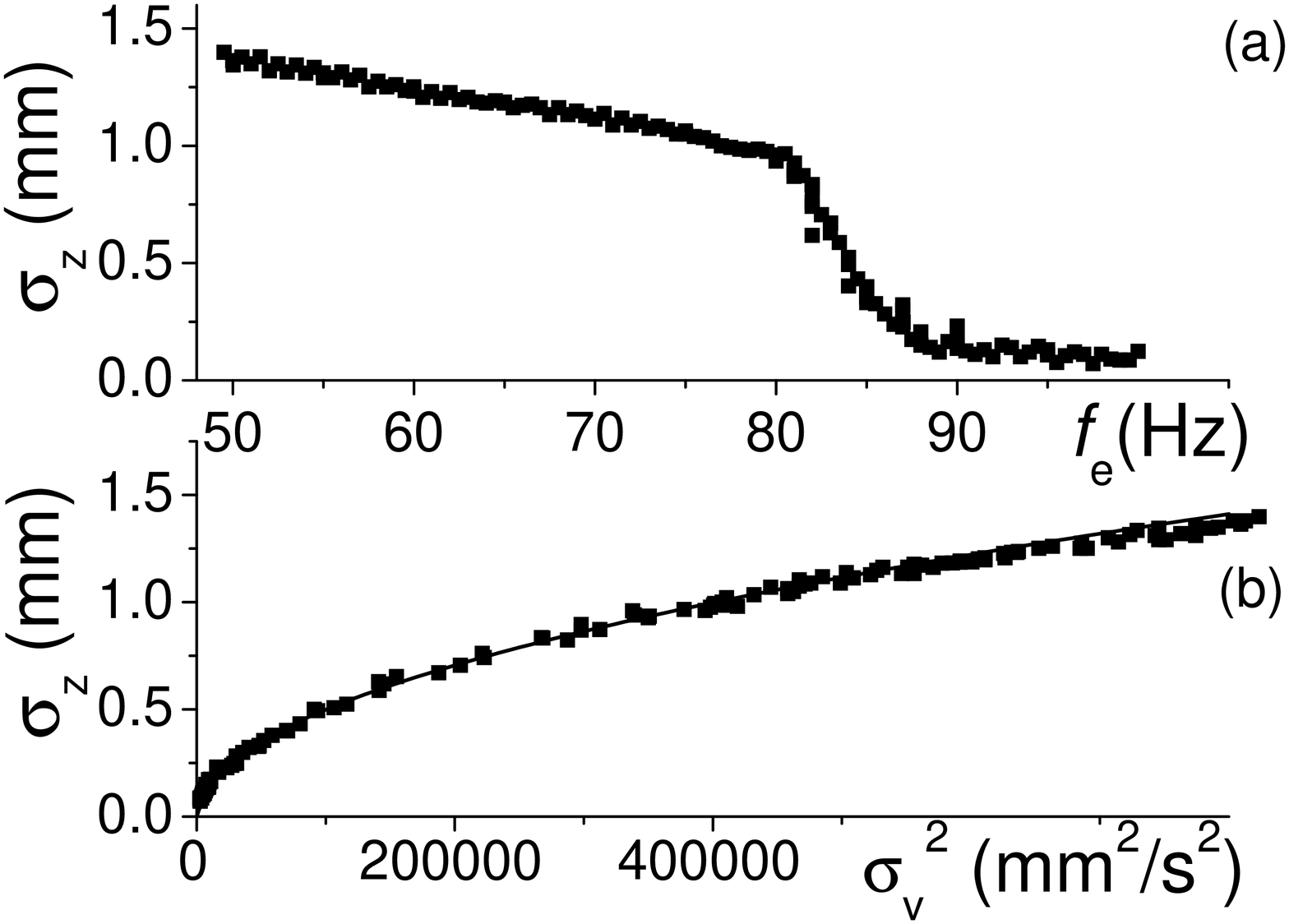}\\
  \caption{(a). Evaporation transition for the experiment of Fig.1. $\sigma _{z}$
as a function of the external driving frequency $f_{e}$. (b)
$\sigma _{z}$ vs. $\sigma _{V}^{2}$ and the corresponding
theoretical fittings. The $f_{e}$ was from
$50-100$Hz.}\label{FIG2:}
\end{figure}
Beyond {\it the} Gibbs ensembles, it should be possible to
conceive a common conceptual framework under the non-equilibrium
thermodynamics \cite {KurchanInOut} that involves jammed states,
as described by Edwards ensembles \cite{Edwards,MakseJamm}. From a
granular gas it is possible to get either a jammed state, for
example by reducing or quenching $\Gamma $ at a given $f_{e}$
\cite{D´AnnaJamm}, or a crystal by slowly scanning up $f_{e}$ in a
fixed accelerated experimental set-up (cooling down). In fact it
is possible to see how a non-Brownian gas is condensed into a {\it
fcc} crystal \cite{FCM04}.

The non-equilibrium GM temperature is a controversial quantity presenting
anisotropy and inhomogeneous characteristics, and its behavior as a function
of the external pumping of energy is yet an unsolved question. However we
have laid out an experimental set-up that makes it possible to measure the
``global'' temperature as the Lagrange parameter that appears in \cite
{PRL9510}.

Since energy is constantly being added to the system a nonequilibrium steady
state (s.s.) mass profile for the GM bed is reached. We report here, upon
measuring the evaporation transition line in a weakly vibrated GM medium,
the dramatic consequences of the exclusion of volume principle, for a
typical non-equilibrium GM system. It is important to remark that below this
critical point the GM behaves as a crystal where the particles move in
``thermally'' activated harmonic oscillations, thus in this regime a
Brownian description is suitable, and this fact was already reported by \cite
{D'AnnaBrownian}. In a previous work \cite{FCM04} we have studied the
spectrum of vibrated GM under gravity, and shown that in the weakly excited
regime the dynamics of these evaporated particles precludes describing them
as Brownian. They show (over 2 decades) a power law spectrum $S_{z}(f)\sim
1/f^{\upsilon }$ with an anomalous exponent $\nu $ ($2\leq \nu <5$) that
depends on the excited collective movement of the grains. This fact leads us
to the conclusion that a description of the present chaotic cooperative
dissipative dynamics of the GM particles, must be done in terms of
non-Markov particles representing the collective effects of the excited GM%
\cite{libro,BuC04}. Recent reports concerning the non-Brownian behavior of
the velocity fluctuations can also be found in resistivity experiments which
show that the electric noise has interesting properties of scale invariance
and of intermittency which arise from thermal expansion locally creating or
destroying electrical contacts \cite{Falcon1}. In addition, associated
anomalous transport properties have been reported as non-Brownian motion
\cite{Falcon2}.

Following Hayakawa and Hong \cite{HHo97} we have introduced an approach to
understand the behavior of a weakly excited GM \cite{PRL9510}. Our $3-$%
dimension $N-$particles experimental conditions allow us to consider a
system of $n-$rows in a cylindrical container as a 1 dimensional degenerate
Fermi-like system. Our laser facility was set up to investigate the
occupation dynamics at the fluidized gap that appears at the top of the $n$%
-rows GM bed. In Fig. 1(c) a typical $z(t)$ is shown. This measurement
corresponds to the beginning of evaporation, where the spectrum exponent is $%
\upsilon =3$, thus we shown that the velocity fluctuation does not
correspond to the one that could be obtained from a Brownian particle \cite
{FCM04}.

By focusing on the configurational properties of an excluded volume theory
the s.s. mass profile can be understood in terms of a configurational
maximum principle assumption. Excluded volume interactions of the GM do not
allow two grains to occupy the same state of gravitational energy. Following
Landau's approach to study a non-equilibrium system, the maximization of $%
S=\ln \prod_{i}\left[ \Omega !/N_{i}!(\Omega -N_{i})!\right] $ yields that
the density profile is $\phi (\epsilon )=\left[ 1+Q\exp (\beta \epsilon
)\right] ^{-1}\left. \Omega \right/ mgD$ where $\Omega $ is the degeneracy
(number of boxes in which to put the maximum amount of balls in a constant
gravitational layer); $\beta $ is a Lagrange multiplier parameter, $%
Q^{-1}=\exp (\beta \mu )$ the fugacity, and $\epsilon =mgDs$ with $%
s=0,1,2,3\cdots .$ Introducing the normalization we get the relation between
the {\it chemical potential} $\mu $ and the {\it temperature} $\beta ^{-1}$,
i.e., $\exp (N\beta mgD/\Omega )=1+\exp (\beta \mu )$. The {\it zero-point}
``chemical potential'' is $\mu _{0}=mgDN/\Omega ,$ so $N/\Omega $ is the
number of balls in an elementary {\it column} of diameter $D$. It is trivial
to understand that without pumping of energy the c.m. is characterized by $%
z_{c.m.}=\mu _{0}/2mg\equiv h/2$. For a weakly vibrated GM we have shown
that $\beta $ is a non-trivial function of the $\sigma _{V}^{2}$ \cite
{PRL9510} (if the particles were Brownons the relation should be $\mid \bar{v%
}\mid \propto \beta ^{-1/2}$).

Shaking gently a box of GM is an example of a nonequilibrium system that can
be characterized by a global dissipative parameter $\delta ,$ which in fact
is given as the ratio of {\it the} two important energies of the system,

\begin{equation}
\delta =\left. 2\frac{mg\Delta z_{c.m.}}{\mu _{0}}\right/ \left( \frac{%
A\omega }{\sqrt{gD}}\right) ^{2}.  \label{Mágica2}
\end{equation}

Since the motion of these {\it mFp} can be studied in terms of the
cumulative probability $\left. \phi (\epsilon =mgz)\right/ N,\ z=Ds$ we have
shown that at low temperature, $\beta \mu _{0}\gg 1,$ we get \cite{PRL9510}

\begin{equation}
\sigma _{z}\simeq \left. 2\Theta \right/ mg\beta .  \label{Beta}
\end{equation}
where $\Theta =\ln (1+\sqrt{1-1/\sqrt{e}})+\frac{1}{4}.$ This expression
gives the amplitude dispersion $\sigma _{z}=\sqrt{\left\langle
z(t)^{2}\right\rangle -\left\langle z(t)\right\rangle ^{2}}=\sigma
_{\epsilon }/mg$ as a function of $\beta .$ The connection between the
fluctuation in the velocity $\sigma _{V}^{2}=$ $\left\langle
V(t)^{2}\right\rangle -\left\langle V(t)\right\rangle ^{2}$ and $\beta $ is
given by our generalized equipartition law: $\frac{1}{2}m^{*}\sigma _{V}^{2}=%
\frac{1}{2}\frac{\Delta N}{N}\beta ^{-1}$, where $m^{*}/m=\delta $ and $%
\left. \Delta N\right/ N$ is a relative factor that counts the ``active''
{\it mFp} in the fluidized gap. From this we get

\begin{equation}
\frac{1}{\beta }\left( \frac{1}{\beta \mu _{0}}\ln \left[ 2e^{\beta \mu
_{0}}-1\right] -1\right) =\delta m\sigma _{V}^{2},  \label{eq000}
\end{equation}
and this formula connects $\sigma _{V}^{2}$, the dissipative parameter $%
\delta $ and $\beta $. So using (\ref{Mágica2}) we relate the collective
response of the system: {\it the expansion} of the c.m. $\Delta
z_{c.m.}=U/N-\mu _{0}/2$, where $U=\int_{0}^{\infty }$ $\epsilon \phi
(\epsilon )\,d\epsilon $, with the variance $\sigma _{V}^{2}$ and $f_{e}.$

Our experiments show that the fluidized gap reveals an inelastic gas heated
in a non-uniform way. Now let us understand this gas of {\it mFp} by
presenting an effective kinetic theory. The collision frequency between {\it %
mFp} is given by $\tau ^{-1}=\bar{V}\sigma _{0}n$, where $\sigma _{0}=\pi
D^{2}$ is the total scattering cross section between two hard spheres, $\bar{%
V}$ their mean relative speed, and $n$ is the mean number of these {\it mFp}
per unit volume. The {\it mean free path} is $L=\tau \bar{v}.$ Proposing the
relative speed $\bar{V}=c_{1}\sigma _{V}$ and the mean speed $\bar{v}%
=c_{2}\sigma _{V}$ (actually $\bar{V}$ should be somewhat larger than $\bar{v%
}$) and using that the mean number of ``active'' {\it mFp} per unit of
volume is given by

\begin{equation}
n=\frac{4c_{3}\Delta N}{\pi D_{c}^{2}2\sigma _{z}},  \label{mFp}
\end{equation}
and assuming all the constants $c_{i}={\cal O}(1)$ we get for $\beta \mu
_{0}\gg 1$

\begin{equation}
L\sim \frac{1}{2}\left( \frac{D_{c}}{D}\right) ^{2}\frac{\sigma _{z}}{\Delta
N},  \label{caminito}
\end{equation}
$L$ calculated from the experimental data $\left\{ \sigma _{z},\Delta
N\left( \sigma _{V}\right) \right\} $ is depicted in Fig. 3(a) as a function
of $f_{e}$, in addition in Fig. 3(b) we show $\beta $ against $f_{e}$ to
note the evaporation transition point. From the linear fitting of Fig. 3(a)
we obtain $L\sim 2.28$mm, this value is indeed only an estimation since we
don't know the values of the constants $c_{i}$.

With a video camera microscope we took photographs of the surface of the
GM-bed from the top of the container and we found that time average pictures
show that $L\sim D$, this result agrees with the measurements of correlation
in the gas phase \cite{Olafsen}. By estimating the degeneration $\Omega $ in
the expression (\ref{caminito}) from $\pi \left( D_{c}/2\right) ^{2}\sim \
\Omega D^{2}$, we get $L\sim 4\Theta D/\pi \ln 2\sim 1.35D$. This important
result shows that a ``jumping model'' can therefore be considered to build
up a kinetic transport theory; thus allowing to calculate the VDF for the
{\it mFp} in the fluidized gap.
\begin{figure}
  \includegraphics[width=5.68cm,angle=-90]{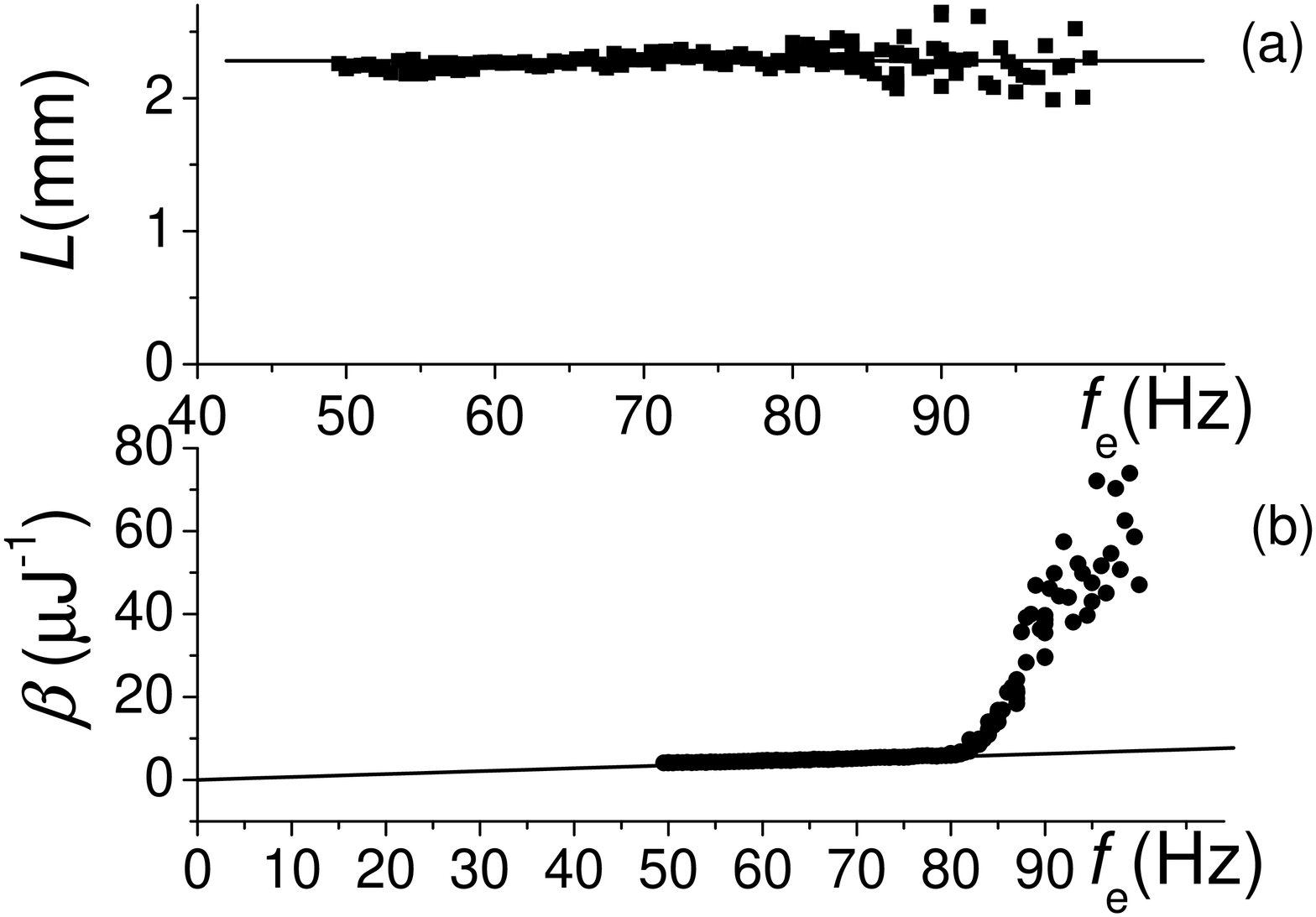}\\
  \caption{(a). Mean-free-path $L$, and (b) the Lagrange parameter $\beta $ vs.
frequency $f_{e}$.}\label{FIG3:}
\end{figure}
We want to emphasize that Fig. 3(a) supports the application of a {\it %
jumping model} to study the kinetics of this non-Maxwellian gas because $%
L\sim D$, i.e., we shall associate a jump probability distribution $P(r)$
(proportional to $\psi (\mu _{0}+\epsilon )\left| \frac{d\epsilon }{dr}%
\right| ,$ where $r=\epsilon /mg>0,$ \cite{PRL9510}) with the transport
mechanism of a set of independent particles in a background of a well
defined global temperature $\beta ^{-1}$. A crucial ingredient for
understanding the dynamics of the GM would be to be able to build up a
statistical mechanics for the construction of the VDF of the driven GM
system. A less ambitious program is to have a model for the kinetic
transport of the {\it mFp} and that is what we are going to present.

Supposing that during the time between collisions the velocity $V$ of each
particle is essentially constant, a jump of length $r$ is performed with
velocity $V=r/\tau >0;$ then defining the kinetic energy as $K=mV^{2}/2$ and
using the transformation of random variables we get $P(K)dK=P(r)dr$. This
``jumping model'' allows us to arrive at a closed expression for the density
of the kinetic energy in terms of the $\beta $ of the s.s. of the
configurational entropy, i.e.,

\begin{equation}
P(K)\propto g\tau \sqrt{\frac{m}{2K}}\ \psi (\mu _{0}+g\tau \sqrt{2mK});\
\int_{0}^{\infty }P(K)\ dK=1.  \label{PK}
\end{equation}
Using the mass profile we get $\psi (\epsilon )=-d\phi /d\epsilon $,
therefore from (\ref{PK}) we obtain an analytic formula for the kinetic
distribution function (KDF) $P(K)$, which predicts at medium and high energy
ranges a larger population than the Maxwellian. Note that in the asymptotic
behavior $\epsilon \gg \mu _{0},$ $\psi (\epsilon )\rightarrow \beta
e^{-\beta \epsilon }$ then KDF for high energy goes like $\sim g\tau \sqrt{%
m/K}\ e^{-g\tau \beta \sqrt{2mK}}$ which is non-Maxwellian.

An important program is to perform a suitable experiment to be able to
determine the functional form of the KDF, and this is what we are going to
present in the rest of the paper. While in the first experiment we scan in $%
f_{e}$ at a given $\Gamma $ to investigate the behavior of $\beta $, in a
second experiment we carry out an isothermal-like measurement during a long
enough time to get the KDF\ and the corresponding $\beta $.

From the measurement of $z(t)$ we were able to calculate the time dependence
of the kinetic energy flux in the measurement of a $1$-dimensional window,
from which it is possible to obtain the KDF. At a given $f_{e}$ and $\Gamma $
we measure $5000$ windows of $2$ seconds to get a total of $25$ millions of
events from which we build up the s.s. KDF. In Fig. 4 we fit our
experimental data by using the formulae (\ref{PK}); the experiment was
carried out for fixed $f_{e}=80$Hz and $\Gamma =10$. Medium and high energy
{\it particles}, which are at the surface of the vibrated GM bed, are well
described by formulae (\ref{PK}), while at low energy the behavior is
Maxwellian. Also we fit the whole experimental results with a Maxwellian
distribution in kinetic energy $\Pi (K)=\exp \left( -K\beta \right) /\sqrt{%
\pi K/\beta }$. As well as with a stretched exponential for the
velocity distribution: $\Pi (V)\sim \exp \left[ -\left(
V/V_{0}\right) ^{\alpha }\right] $, with $\alpha =3/2$, where
$V_{0}$ is the thermal {\it rms} velocity\cite{Ernst}; this last
fitting shows that the agreement is also good. Since Olafsen et
al. reported the stretched exponential law for the VDF\ of the gas
phase of a vibrated monolayer GM\cite{Olafsen}, several
research groups have been measuring this law for granular gases \cite{Menon}%
. A delicate obstacle for the analytical solution of this problem is the
supply of energy to the granular bed which {\it compensates} the dissipation
caused by inelastic collisions \cite{Ernst}, an aspect which was clarified
by van Zon el at. using simulations of molecular dynamics \cite{MacKintosh}.
For granular gases there is indeed a family of distributions, with apparent
exponents $\alpha <2$ which are governed by the ratio between the numbers of
heating events and inelastic collisions \cite{MacKintosh}. Either under
uniform heating or boundary heating of the GM, if the dissipative collisions
dominate the heating process produces a liquid-like cluster surrounded by a
gas phase and the VDF is strongly non-Gaussian. Also after Olafsen et al.
the conclusions of several authors keep clear the conceptual relationship
between this high energy tail behavior, the exclusion of volume and the
clustering in granular gases. Our description adds a new picture for
clarifying this delicate point, the $\delta $ parameter defined in equation (%
\ref{Mágica2}) and the generalization of the equipartition law (\ref{eq000})
permit us to tackle this obstacle, to give the KDF in terms of $\beta $. Our
approach was obtained from the physical parameters: $\{f_{e},\Gamma ,\mu
_{0},D,N,\Omega \}$ and by using the free parameter $\delta $ from the
experiment described in Fig. 2 and 3. The fitting for the curve in Fig. 4
gives $\beta =(1.7\pm 0.5)(\mu $J$)^{-1}$ and for the corresponding value
from Fig. 3(b) we get ($5\pm 1)(\mu $J$)^{-1}$, which in fact is a good
agreement considering that in the first experiment we are scanning in $f_{e}$
at constant rate, and in the second experiment (time average) both $f_{e}$
and $\Gamma $ are fixed. The high energy tail corresponds to the beads near
to the surface of the GM bed where the jumping combined with collisions
dominates the process.
\begin{figure}
  \includegraphics[width=5.68cm,angle=-90]{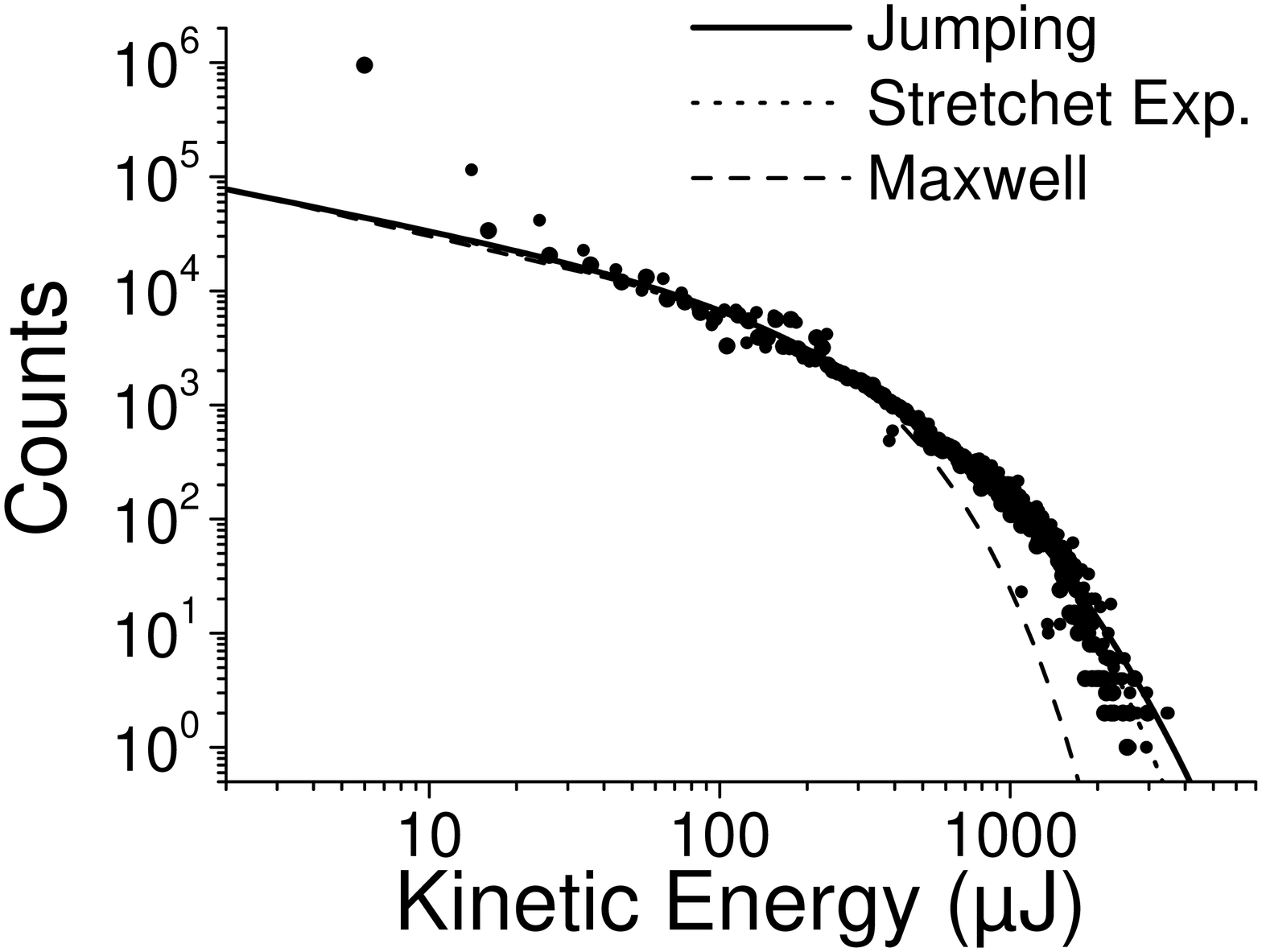}\\
  \caption{Experimental kinetic energy distribution function (KDF) and the
corresponding fittings according to our ``jumping'' kinetic model,
the Maxwellian and the stretched exponential cases.}\label{FIG4:}
\end{figure}

A point of special interest is that the {\it packing fraction} for
this fluidized gap is constant for the whole range of low
temperatures we used, which can be estimated as:

\begin{equation}
p_{f}=\frac{2}{3}\left( \frac{D}{D_{c}}\right) ^{2}\frac{D\cdot \Delta N}{%
\sigma _{z}}\sim \frac{\pi }{12}\frac{\ln 2}{\Theta }=0.25.
\label{packingfactor}
\end{equation}

At constant cooling down by increasing $f_{e}$ at constant $\Gamma $, this
lattice-gas condenses into an $fcc$ crystal, and vice-versa from the $fcc$
lattice under constant heating rate the expansion of the c.m. entails the
expansion of this $fcc$. This expansion under gravity creates fluctuations
in the number of particles $n(\epsilon )$ around $\epsilon =\mu _{0}$, which
induces the transport of matter at the fluidized-gap. The time integration
of such fluctuations (as we did during $3$hs. in \cite{PRL9510}) reveals
that the fluidized-gap has a structure with a Fermi profile. Thus we
conclude that for our experimental conditions the gaseous gap is indeed a
lattice-gas ({\it fcc} where half of sites are occupied) with $p_{f}$
according to equation (\ref{packingfactor}). Nevertheless, the picture we
get for this lattice-gas should be understood as a ``jammed state of
motion'', since the $p_{f}$ does not change for the range of low
temperatures we measured.

The important experiment of D'Anna et al. \cite{D'AnnaBrownian} shows that a
vibrated granular bed exhibits a formal analogy with a {\it thermal bath},
where it is possible to apply a fluctuation-dissipation theorem. In that
experiment a harmonic oscillator of frequency $\omega _{0}$ (the torsion
pendulum) is coupled to the Brownian particles in the vibrated granular bed.
This thermal bath is generated by driving the GM in a frequency band between
$300-900$Hz at a given $\Gamma $ (from $1$ to $11.6g$); in fact, out of
equilibrium this thermal bath is thought within the frame in \cite
{KurchanInOut}. The D'Anna's torsion pendulum is this thermometer coupled to
the Brownian ensemble, the driven fluctuation pumps energy while its
interaction with the same Brownons provokes the dissipation of its energy.
Nevertheless if evaporation starts the viscosity of the thermal-like bath
should therefore be related to the exclusion of volume, since the jumping of
the particles is conditioned by the availability of free volume.

In a system in equilibrium and in D'Anna's experiment, equipartition means
that $K$ must be $k_{B}T/2$ and by measuring long enough they get a
temperature for the thermal bath (or granular bed), the range of
temperatures tested were lower than $0.1\mu $J. Our laser-thermometer
explores beyond the Markovian range of the GM bed, which happens when the
jumping events start (evaporation) and the fluidized gap appears. Our
thermometer is a test particle of the GM bed which occupies the position
tested by the laser.

Measurement of $\beta ^{-1}$ as a function of $f_{e}$ exhibits at $10$
PKelvin the critical point that characterizes the evaporation transition.
For $\beta ^{-1}<10$PKelvin ($\beta ^{-1}<0.1\mu $J) the system is condensed
and the motion corresponds to Brownian oscillators. If evaporation starts
the following picture is possible: when our ensemble of {\it mFp} is
perturbed, the energy is dissipated by the mechanisms of jumping (under
gravity) and random inelastic collisions between the {\it mFp}, but
conditioned to the fact that these events are self-organized in time,
keeping constant the $p_{f}$ of the fluidized gap. The relation between the
transport and the number of ``active'' particles $\Delta N$ (at the Fermi
sea) links the two Lagrange parameters: $\beta $ and $\mu $ (compactivity),
and therefore describes the existence of a lattice-gas. The coupling between
$\beta ^{-1}$ and $\mu $ tells us that the important rule, for the ensemble
of {\it mFp}, is to dissipate energy keeping constant the lattice structure.
We remark that the difference with the well known Markovian motion is that
Brownons can dissipate energy through viscosity independently of any gas
structure.

Our approach focuses on the potential energy which is a function of the
configurational entropy and the relation with the kinetic energy per active
particle at the gap. In other words, while in the Brownian situation we can
describe the system in terms of energy fluctuations, viscosity and
temperature, beyond the Markovian approximation, the important events are
the fluctuations in the number of particles. Then we are indeed reporting on
the existence of a {\it compensation} relation between the fluctuations in
the number of {\it mFp }at the gap (or the matter transport), the Fermi
profile held by the $\delta $ dissipative parameter, and the global
temperature $\beta ^{-1}$. A fluctuation in the number of {\it mFp} in the
gap will be compensated in time in such a way that the lattice-gas keeps $%
p_{f}$ constant: this is a fluctuation-compensation relation. Therefore a
stochastic formalism is still valid but the equations should describe a
fluid where the complexity is the description of the lattice-gas, the
jumping phenomena and collisions in a space framed by the exclusion of
volume.

This approach agrees with the discussion in \cite{KurchanInOut} related to
the idea that systems far from equilibrium can be described with more than
one temperature. Such temperatures are related to time constants of the
different configurational structures. For a glass, the time for relaxation
to equilibrium is extremely long, and it is possible to describe this system
by using more than one thermometer with different time constants. In GM\
experiments, the structure of the gap is only measurable if we wait enough
time. This structure is related to transport of mass at the gap, as is
revealed by our generalized equipartition law. The characteristic time we
must wait to measure $\beta $ in this system is the time we need to
integrate the mass profile. However unlike glasses, GM vibrated under
gravity is far away of equilibrium but stationary: the system exists while
the $\delta $ parameter is held at a fixed value, and the
fluctuation-compensation should play a role stabilizing the steady state,
i.e., the precise {\it balance} of fluctuations permits us to know the
global temperature without waiting too long.

{\it Acknowledgments}. JEF thanks the von Humboldt Foundation and the
DFG-Graduiertenkolleg 1276/1, Prof. C. Wagner and Dr. G. C. Rodriguez. MOC
thanks Prof. V. Gr\"{u}nfeld for the English revision, and grants from
SECTyP, Uni. Nac. Cuyo; and PIP 5063 (2005) CONICET, Argentina.


\begin{references}
\bibitem{PRL9510}  Fiscina J.E. \& C\'{a}ceres M.O. Fermi-like behavior of
weakly vibrated granular matter. Phys.Rev.Lett. 95, 108003 (2005).

\bibitem{KurchanInOut}  Kurchan J. In and out of equilibrium. Nature 453,
222 (2005).

\bibitem{Edwards}  S.F. Edwards. Granular Matter: An Interdisciplinary
approach. edited by A. Metha, Springer Verlag, New York (1994) and
references therein.

\bibitem{MakseJamm}  Makse H.A. \& Kurchan J. Testing the thermodynamic
approach to granular matter with a numerical model of a decisive experiment
Nature 415, 614 (2002).

\bibitem{D´AnnaJamm}  D'Anna G. \& Gremaud G. Nature 413, 407(2001).

\bibitem{FCM04}  Fiscina J.E., C\'{a}ceres M.O. \& M\"{u}cklich F. On the
spectrum behaviour of vibrated granular matter J. Phys.: Condens. Matter{\bf %
\ 17}, S1237 (2005).

\bibitem{D'AnnaBrownian}  D'Anna G., Mayor P., Barrat A., Loretto V. \& Nori
F. Observing brownian motion in vibration-fluidized granular matter. Nature
424, 909 (2003).

\bibitem{libro}  C\'{a}ceres M.O. Estadistica de no equilibrio y medios
desordenados, {\it in Spanish} (Revert\'{e} S.A., Barcelona, 2003).

\bibitem{BuC04}  Budini A.A. \& C\'{a}ceres M.O. Functional characterization
of generalized Langevin equations. J. Phys. {\bf A 37}, 5959, (2004).

\bibitem{Falcon1}  Falcon M.E., Castaing B. \& Creyssels M. Nonlinear
electrical conductivity in a 1D granular medium. Eur. Phys. J B. 38, 475
(2004).

\bibitem{Falcon2}  Falcon E., Castaing B. \& Laroche C. Turbulent electrical
transport in copper powders. Eurphys. Lett. 65, 186 (2004).

\bibitem{HHo97}  Hayakawa H. \& Hong D. Thermodynamic theory of weakly
excited granular systems. Phys. Rev. Lett. {\bf 78}, 2764 (1997).

\bibitem{Olafsen}  Olafsen J. S. \& Urbach J.S. Clustering, order, and
collapse in a driven granular monolayer. Phys. Rev. Lett. {\bf 81},
4369,(1998).

\bibitem{Ernst}  Barrat A., Trizac E. \& Ernst M.H. Granular gases: dynamics
and collective effects. J. Phys.: Condens. Matter {\bf 17, }S2429 (2005);
van Noije T.P.C \& Ernst M.H. Velocity distributions in homogeneous granular
fluids: the free and the heated case. Granular Matter{\bf \ 1}, 57 (1998).

\bibitem{Menon}  Rouyer F. \& Menon N. Velocity fluctuations in a
homogeneous 2D granular gas in steady state. Phys. Rev. Lett. 85, 3676
(2000). Feitosa K. \& Menon N. Fluidized granular medium as an Instance of
the fluctuation theorem. Phys. Rev. Lett. 92, 164301 (2004).

\bibitem{MacKintosh}  van Zon J. S. \& MacKintosh F. C. Velocity
distributions in dissipative granular gases. Phys. Rev. Lett. 93 038001
(2004).
\end{references}
\end{document}